\documentclass[a4paper,twocolumn,11pt]{scrartcl}
\usepackage[top=2cm,bottom=3cm,left=2cm,right=2cm]{geometry}
\RedeclareSectionCommand[font=\centering]{section}

\pdfoutput=1

\usepackage{array}
\usepackage{soul}
\usepackage{xcolor}
\usepackage{mathtools}
\usepackage{amsmath,amsfonts,amssymb}
\usepackage{graphicx,bm}
\usepackage{booktabs}
\usepackage[colorlinks,linkcolor=blue,urlcolor=blue,citecolor=blue,anchorcolor=blue]{hyperref}
\usepackage{slashed}
\usepackage{cite}
\usepackage{ulem}

\usepackage[small,bf,format=plain]{caption}
\DeclareMathOperator{\diff}{\text{d}}
\DeclareMathOperator{\ci}{\text{i}}
\newcommand\brabar{\scalebox{.3}{(}\raisebox{-1.7pt}{--}\scalebox{.3}{)}}

\begin{document}

\twocolumn[
\begin{center}
{\titlefont \Large Bremsstrahlung from Neutrino Scattering via Magnetic Dipole Moments \par}
\vspace{3ex}
{Konstantin Asteriadis$^{\, a,\,}$\footnotemark, Alejandro Quiroga Trivi\~no$^{\, b,\, c,\,}$\footnotemark \ and Martin Spinrath$^{\, b,\, c,\, d,\,}$\footnotemark\par}
\vspace{2ex}
{\textit{\small $^{a}$High Energy Theory Group, Physics Department,\\Brookhaven National Laboratory, Upton, NY 11973, USA}\par}
\vspace*{0.2cm}
{\textit{\small $^{b}$Department of Physics, National Tsing Hua University, Hsinchu, 30013, Taiwan}\par}
\vspace*{0.2cm}
{\textit{\small $^{c}$Physics Division, National Center for Theoretical Sciences, Taipei 10617, Taiwan}\par}
\vspace*{0.2cm}
{\textit{\small $^{d}$Center for Theory and Computation, National Tsing Hua University, Hsinchu, 30013, Taiwan}\par}
\vspace{3ex}
{\titlefont Abstract\par}
\vspace{2ex}
\parbox{0.9\linewidth}{
{\small In this paper we discuss  
bremsstrahlung induced by neutrino scattering.
This process should exist since neutrinos are expected to couple to photons
via magnetic dipole and transition moments. These moments are loop-induced and
tiny in the Standard Model with neutrino masses
but could be significantly enhanced in extended theories.
As concrete example we study the scattering of the two largest neutrino
fluxes on earth, solar neutrinos and Cosmic Neutrino Background (CNB).
It is tempting to consider this as a potential signature for CNB
searches but it turns out that the signal is extremely small
and unlikely to be observed. 
}}
\par\vspace{10pt}
\end{center}
]

\footnotetext[1]{\href{mailto:kasteriad@bnl.gov}{kasteriad@bnl.gov}}
\footnotetext[2]{\href{mailto:alejandro.quiroga@gapp.nthu.edu.tw}{alejandro.quiroga@gapp.nthu.edu.tw}}
\footnotetext[3]{\href{mailto:spinrath@phys.nthu.edu.tw}{spinrath@phys.nthu.edu.tw}}

\section{Introduction}

Neutrinos remain to be among the most fascinating particles
due to their importance in particle physics, nuclear physics,
astrophysics and cosmology. For instance, they are the only known particles
with confirmed properties beyond the Standard Model of
particle physics.

In cosmology one of their interesting aspects is that they
form their own background, the Cosmic Neutrino Background (CNB),
which is the equivalent to the Cosmic Microwave Background.
The direct observation of the CNB in
a laboratory remains one of the great challenges of experimental
particle cosmology. Given the tiny cross sections and energies
of CNB neutrinos it was even called an ``apparently impossible
experiment''~\cite{Melissinos:1999ew}.
Some recent proposals include detecting the tiny force induced
by the CNB ``wind'' using current gravitational wave
detector technology~\cite{Domcke:2017aqj,Shergold:2021evs}, resonant scattering
against ultra-high energetic cosmic 
neutrinos~\cite{Brdar:2022kpu}, cosmic birefringence induced by the CNB
\cite{Mohammadi:2021xoh} and the absorption of CNB neutrinos on
tritium~\cite{PTOLEMY:2018jst, Betts:2013uya}. The last is probably the most
promising proposal at this time, see, e.g.,~Refs.~\cite{Gelmini:2004hg,Ringwald:2009bg,Vogel:2015vfa,Bauer:2022lri}
for more comprehensive reviews.

The original motivation for this work was to study an alternative
method for its detection.

The idea is to look for a signature of the
scattering between the two largest natural neutrino fluxes on earth: 
solar neutrinos and CNB neutrinos.
Since the scattered neutrinos would still be hard to detect, we consider
an additional bremsstrahlung photon in the final state that is
comparatively easy to detect and that can be produced at any energy.
To our knowledge, this is the first time this process was calculated.
In other cases neutrinos were considered as the final states of the
bremsstrahlung itself, i.e., \cite{Hannestad:1997gc,Bhattacharyya:2005zb} and
references therein. 

In the Standard Model of particle physics (SM) including Dirac neutrino masses neutrinos couple to photons via loop induced magnetic dipole moments.
The neutrino magnetic moment is tiny and so is the cross section for the considered process.
For Majorana neutrinos, however,  the magnetic moment is exactly zero.
In this case, the scattering cross section is still non-zero if one considers transition 
magnetic moments, see, e.g.~Ref.~\cite{Czakon:1998rf}, and we expect the cross sections to be similar 
to the (simpler) Dirac case.
In any case, this process should exist even in the SM and lead to a tiny photon 
flux throughout the universe.
Nevertheless, as we will see the event rate for this process is
extremely small and an observation of this photon flux seems rather unlikely
even under optimistic assumptions. Although there might be extreme environments
with large fluxes of high-energetic neutrinos where the considered process
might become interesting.

\section[Bremsstrahlung from neutrino-neutrino scattering]{Bremsstrahlung from\\ neutrino-neutrino scattering}

We study the processes 
\begin{equation}
\label{eq:processes}
\nu_\odot \; + \stackrel{\brabar}{\nu}_{\text{CNB}}  \to  \nu \; +\stackrel{\brabar}{\nu} + \; \gamma 
\end{equation}
at leading order, neglecting the exchanged momentum
with respect to the $Z$-boson mass in the propagators, see also the Appendix for more details.
Here $\nu_\odot$ is a solar neutrino and $\nu_{\text{CNB}}$ and $\bar{\nu}_{\text{CNB}}$ 
are relic neutrinos and anti-neutrinos, respectively. In the SM  
and standard cosmology the CNB is expected to consist of neutrinos and anti-neutrinos
to equal parts and we assume neutrinos to be left-helical and anti-neutrinos to be right-helical today, cf.~Ref.~\cite{Long:2014zva}.
The solar neutrinos are assumed to be purely left-chiral and for the
sake of simplicity we will neglect any flavor effects to get an estimate of the rate of these processes.
We will comment later on how a more realistic flavor treatment can modify our results.
To that end, we treat solar and CNB neutrinos to consist of only one flavor.
We set the neutrino mass to be $m_\nu = 0.05$~eV
which is a mass scale compatible with current limits~\cite{ParticleDataGroup:2022pth}.
For that scale the CNB neutrinos
are non-relativistic and we can neglect their velocity in our calculations.
Interestingly though, non-vanishing CNB velocities would lead to corrections
which would, in principle, allow to measure their velocity distribution.

In our setup, the photons couple to the neutrinos via an effective
magnetic dipole moment with the effective Lagrangian~\cite{Giunti:2014ixa}
\begin{align}
	\mathcal{L}_{\text{eff}} = - \ci M_\nu \, \bar{\nu} \, \sigma_{\alpha \beta} \, q^\alpha \nu \, A^\beta \;,
\end{align}
where $\sigma_{\alpha \beta}$ is the anti-symmetric combination
of $\gamma$-matrices, $q^\alpha$ is the momentum carried away by the
photon field $A^\beta$, and $M_\nu$ is the magnetic moment of the neutrino.
In general, $M_\nu$ is a matrix in flavor space. Since we do a one-flavor approximation
$M_\nu$ is just a dimensionful number.
The coupling then reads
\begin{align}
\label{eq:vertex}
	\vcenter{\hbox{\includegraphics[scale=.41,trim=25 0 0 0,clip]{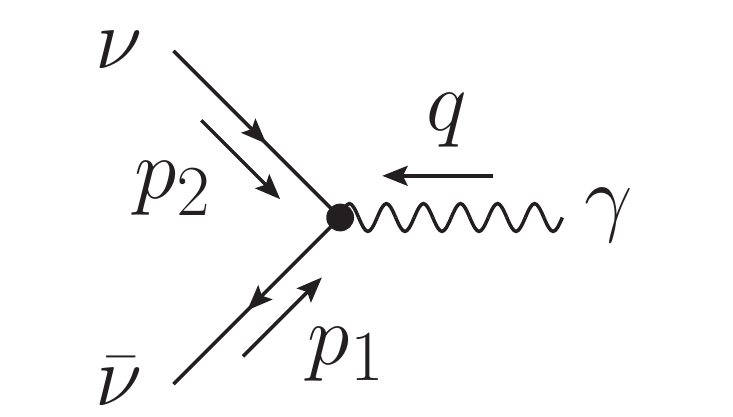}}} \widehat{=} \hspace{7pt}   -\frac{\ci}{2} ( \gamma_\beta \slashed{q} - \slashed{q} \gamma_\beta)   M_\nu \;.
\end{align}
In the SM the coupling~Eq.~\eqref{eq:vertex} occurs for Dirac neutrinos via loops with an effective coupling constant of $M_\nu^\textrm{SM} \lesssim 3.8 \times 10^{-19} \mu_B$~\cite{Giunti:2014ixa}. 
Here we assume it to be additionally enhanced by
some new physics. The parameter $M_\nu$ has been recently constrained by XENONnT to be
$M_\nu < 6.4 \times 10^{-12} \, \mu_B$
at 90\%~CL \cite{XENON:2022ltv}.
In the following, we write $M_\nu = f_M \times 10^{-12} \, \mu_B$ for some $f_M \lesssim 6.4$.
The total rate within our assumptions is proportional to $|f_M|^2$ which makes it easy to rescale
our results to the true value of $M_\nu$.
Note that, for Majorana neutrinos $f_M = 0$, but considering multiple flavors and transition magnetic moments
instead would also imply
the existence of the considered process with expected results similar to the case of Dirac neutrinos.

Neutrinos can also have other electromagnetic moments. 
For instance, they could have a tiny electric charge. However, the current upper bound is so low
that these contributions should be orders of magnitude smaller
(even considering possible enhancements close to infrared singularities).
For this reason, and for simplicity, we will neglect such complications here.

For this set of assumptions, the relation between the
differential cross sections
for CNB neutrinos $\sigma_{\nu\nu}$ and CNB anti-neutrinos
$\sigma_{\nu\bar{\nu}}$ and the differential photon production rate $\textrm{d}R$ can be estimated as
\begin{align}
\begin{split}
\label{eq:Rate}
	\textrm{d}R &= n_{\text{CNB}}  \int \textrm{d}E_\nu \\
	&\hphantom{{}={}}\times\frac{\diff \Phi_{\odot} }{\diff E_\nu} \, \big[\textrm{d}\sigma_{\nu \nu}{(E_\nu)} +  \textrm{d}\sigma_{\nu \bar{\nu}}{(E_\nu)}\big]  \;,
\end{split}
\end{align}
where the local CNB density is chosen to be $n_{\text{CNB}} = f_n \times 168$/cm$^3$.
For $f_n = 1$ this density corresponds to the SM prediction.
The factor $f_n$ parametrizes potential overdensities
which are nevertheless not expected to be very large, see, e.g.\ Ref.~\cite{Ringwald:2004np}.
The scaling factor $f_n$ can also depend on
the neutrino flavor.
Finally, $\diff \Phi_{\odot}/\diff E_\nu$ are the differential solar neutrino
fluxes
which have some energy-dependent flavor dependence in reality due to neutrino oscillations.

In our calculations we consider a target volume of 1~km$^3$, at an earth-like
distance from the sun.
It is considered isolated from the surroundings while CNB and solar neutrinos can still
enter and interact inside the volume.
We calculate how many photons are produced
in it within a year and how their energies and angles are distributed.
The inner surface of the volume could be covered with suitable photon detectors but
we do not want to go into further experimental and technical details which are beyond the scope
of our theoretical work.

Numerical tables for the solar neutrino fluxes are taken from Ref.~\cite{Vitagliano:2019yzm}.
The final state phase space integration is performed numerically
and getting the desired distributions is straightforward.

\section{Results}

\begin{figure}
\centering
\includegraphics[height=205pt]{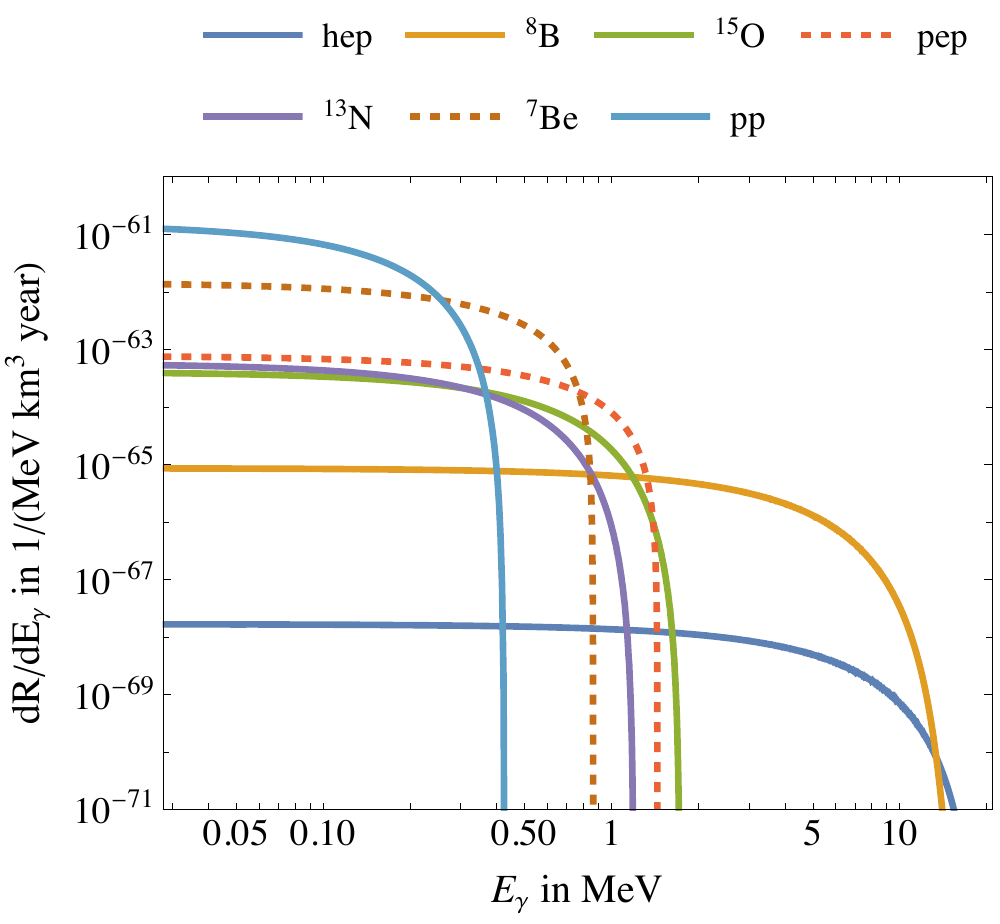}
\caption{\label{fig:dRdEg}
	Energy distributions of the emitted bremsstrahlung photons. Here we have set
	$f_M = f_n =1$ and we separate the different components
	of the solar neutrino flux.
	}
\end{figure}

For numerical results reported in this section, we set $f_M = f_n = 1$
and use $G_F = 1.1663787 \times 10^{-5} \, \textrm{GeV}^{-2}$ for
the Fermi coupling constant and $\mu_B = 296.238 \, \textrm{GeV}^{-1}$ for the Bohr magneton.

In Fig.~\ref{fig:dRdEg} we show the energy distribution
of the emitted bremsstrahlung photons, separating
the different components of the solar neutrino flux.
It follows from this figure that they could be theoretically distinguished from each other.
This is quite a unique
feature which could potentially be used to separate the bremsstrahlung photons
of this process from other potential backgrounds. We also see that the higher energies
of the $^8$B and the hep neutrinos, implying larger cross sections,
cannot compensate for the much larger
flux of the pp neutrinos.
We checked numerically that in the relevant energy range the cross section
grows quadratically with the incoming neutrino
energy to a good approximation.
This increase is much weaker than the flux decrease
for the high energy solar neutrinos.
For that reason
we also do not consider other
naturally occurring neutrino fluxes such as atmospheric neutrinos which
are much smaller than the solar flux~\cite{Vitagliano:2019yzm}.

The energy spectrum can be affected by flavor effects. First of all,
the endpoints of the spectra depend on the actual neutrino masses
which is a tiny effect considering the size of the neutrino masses
compared to the photon energies near the endpoint. Furthermore,
the solar neutrino flux has an energy dependent flavor composition which together
with a non-trivial flavor structure of the dipole moments matrix 
might have a larger effect on, e.g., the shape of the energy distribution but is not
expected to give an overall strong enhancement.
Therefore we chose to keep the discussion of
flavor effects simplistic in this work to get a first estimate of the size of this
process.

\begin{figure}
\centering
\includegraphics[height=205pt]{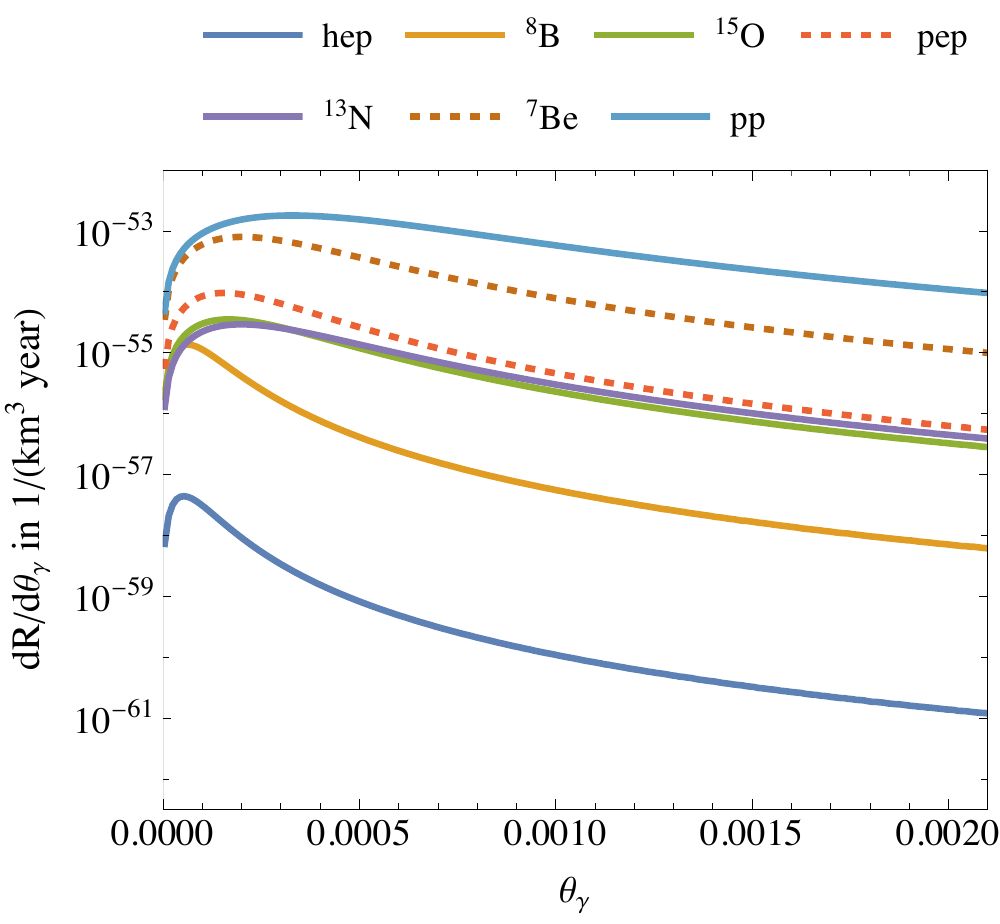}%
\caption{\label{fig:dRdthg}
	Angular distribution of the emitted bremsstrahlung photons.
	The angle is chosen such that $\theta_\gamma = 0$ points away from
	the sun which is treated as a point source. Here we have set
	$f_M = f_n =1$.
	}
\end{figure}

In Fig.~\ref{fig:dRdthg} we show the angular distribution with respect to the
direction towards the center of the sun, assuming the incoming neutrino momenta to be parallel.
As expected the distributions peak at $\theta_\gamma$ very close to zero. 
which is due to the ``fixed target'' nature of our setup
and the very large energies of the incoming neutrinos
compared to the target neutrino mass.
This is another feature which could,
in principle, be used to distinguish a signal from potential background photons.
The width of the angular distribution is similar to the deviations that would be induced by the neutrino production
zones in the sun which have a similar angular diameter
in the sky~\cite{Lin:2022jyv}.
Therefore, a fully realistic treatment 
would widen the distribution somewhat but there would still be 
a very strong directional dependence.

We assumed a detector volume that is shielded from any photons from the surroundings.
Even in this ideal case there should be other photons, for instance, from neutrino decays ($\nu_i \to \nu_j + \gamma$) or Bremsstrahlung
from neutrinos scattering from residual matter in the considered volume
($\nu + X \to \nu + X + \gamma$). However, these processes have different energy and angular distributions than the considered process and are therefore distinguishable in theory.

What becomes apparent from both figures is that the expected rates
are extremely small. In fact, the total expected rate is about $2.1 \times 10^{-56}$
bremsstrahlung photons per year and km$^3$ of target volume assuming no neutrino overdensities
and a neutrino magnetic dipole moment slightly below the current bound, to be precise $f_M = f_n = 1$. That makes the discovery prospects of the CNB using this signature rather
unlikely as one would need a neutrino beam with significantly higher
energies, flux and/or observed volume to get a reasonable rate.
Even if we consider a target volume as large as the earth covered with photon detector
the rate is still
of the order of $10^{-42}$ photons per year. Without the enhancement
by new physics this number drops by an additional fourteen orders.

The dependence of the rate on the model parameters $f_M$ and $f_n$ is only quadratic and linear, respectively.
Increasing the rate by orders of magnitude would also require increasing these parameters by orders of magnitudes which is not expected neither from experiments nor simulations.

What might be more promising to improve the rate is to increase the neutrino flux.
Given that the biggest neutrino flux on earth are the
CNB neutrinos themselves, self-scattering of CNB neutrinos may be an option.
Such a process with similar kinematics to the process studied above would be a massless neutrino flavor
scattering from a massive one. 
We can provide a rough estimate for this case.
The flux would be roughly a factor 25 larger than the solar neutrino flux.
On the other hand the cross section would drop by a factor
$m_\nu^2/E_\odot^2 \approx 2.5 \times 10^{-15}$ where we used
for the neutrino mass $m_\nu = 0.05$~eV and for the solar neutrino
energy $E_\odot \approx 10^6$~eV.
This estimate is only an upper bound because the neutrino mass is used as the energy scale of the CNB self-scattering process instead of the much smaller kinetic energy.
We conclude from the above numbers that 
this process would be even more rare compared to the one involving
solar neutrinos.
This said, this little thought experiment shows how the rates
for other sources can be estimated as long as the center of mass
energy is below the $Z$-boson mass and no other new physics scenarios are considered.

\section{Conclusions}

In this work we calculated for the first time the rate of bremsstrahlung photons
produced in neutrino scattering. As concrete example we consider the two largest neutrino fluxes
on earth, the solar neutrino flux and CNB neutrinos.
Although we chose for our numerical results
a magnetic dipole moment which is strongly enhanced compared to the SM
and just slightly below the current experimental
bound, the obtained cross section and rate is still tiny and somewhat discouraging
for any experimental efforts in an earthbound laboratory in that direction.
Notwithstanding, this process exists even in the SM and should generate a minuscule photon
flux throughout the universe.
While our original motivation was to study another way to detect the
CNB our proposal fails in that regard and rather serves as
a showcase how difficult that endeavour is.

We also showed how the bremsstrahlung photon rate for other
energetic neutrino sources can be easily estimated assuming a similar
set of assumptions.
Other neutrino beams or sources, non-standard cosmology,
additional new physics contributions, a larger target volume
and a combination thereof
could lead to a rate closer to being measurable.
In particular, if it would not involve CNB neutrinos.
We leave the study of these cases for further investigations.

\section*{Acknowledgments}

We want to thank Jan Tristram Acu\~na for some useful comments.
The research of KA is supported by the United States Department of Energy under 
Grant Contract DE-SC0012704.
AQT and MS are supported by the Ministry of Science and Technology (MOST) of Taiwan 
under grant number MOST 110-2112-M-007-018 and MOST 111-2112-M-007-036.

\appendix

\section*{Appendix: Details on the Cross Section Calculation}
\label{Appendix}

We give here the explicit expressions for the matrix elements of the considered processes.
We consider a one-flavor approximation of Dirac neutrinos and
only neutrino magnetic dipole moments are taken into account. 
We furthermore neglect
the exchanged momentum in the $Z$-boson propagator.

\begin{figure}
\centering
\includegraphics[width=0.9\linewidth]{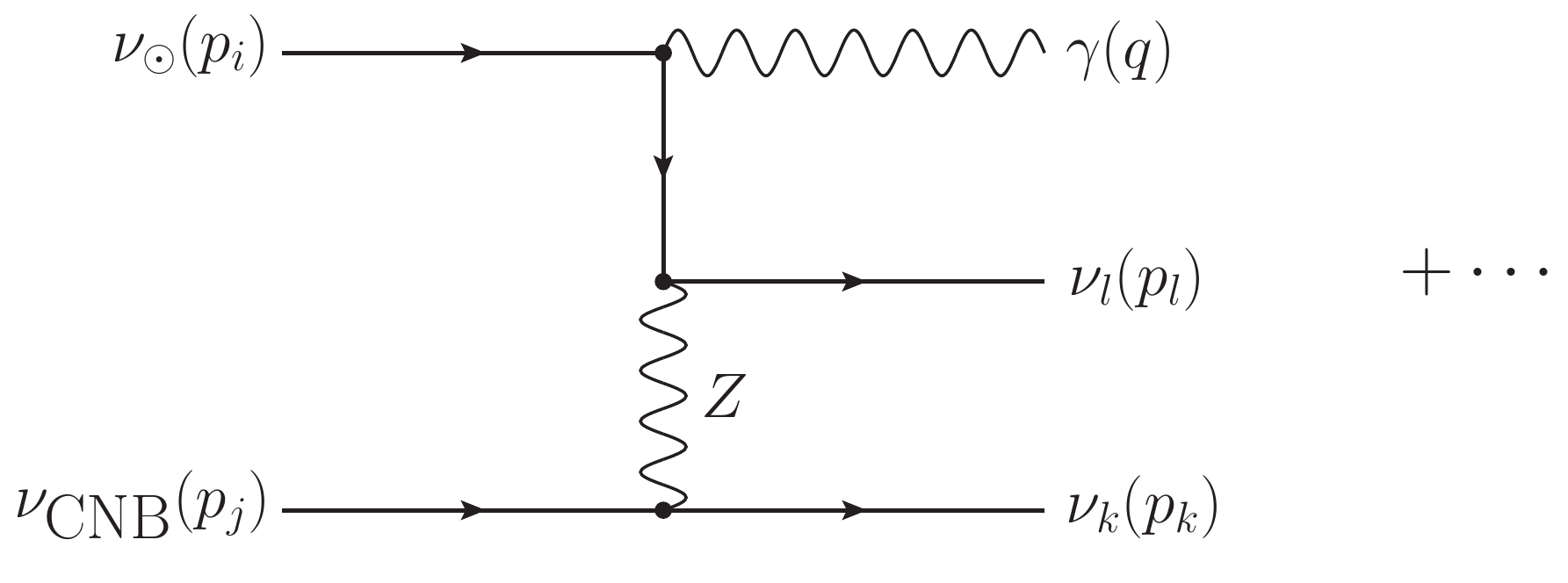}
\caption{\label{fig:FeynmanN}
	Example Feynman diagram for the bremsstrahlung emission from neutrino-neutrino
	scattering. We assume in this paper that the photons couple to neutrinos
	via a magnetic dipole moment, cf.~Eq.~\eqref{eq:vertex}.
	There are seven more diagrams. Three more diagrams where the photon
	couples to the other neutrino lines and four more where the indices
	$k$ and $l$ are interchanged.
	The latter four would not contribute for different neutrino flavors
	in the final state.
	}
\end{figure}

For the process involving only neutrinos, cf.~Fig.~\ref{fig:FeynmanN},
\begin{equation}
	\nu_\odot(p_i) + \nu_{\text{CNB}}(p_j)  \to  \nu (p_k) + \nu(p_l) +  \gamma (q) \;,
\end{equation}
we find the for the spin-averaged, squared amplitude
\begingroup
\allowdisplaybreaks
\begin{align}
&\frac{|\mathcal{M}(\nu \nu \to \nu \nu \gamma)|^2}{128 \, G_F^2 \,  M_\nu^2} =  
6 \, (p_i \cdot p_j) (p_k \cdot p_l) \nonumber\\
&+ m_\nu^2  \biggl[
2 \, (p_i \cdot p_l) + 2 \, (p_i \cdot p_k) - 4 \, (p_i \cdot p_j)
\nonumber\\
&
+ (p_j \cdot p_l) +  (p_j \cdot p_k) - 2 \, (p_k \cdot p_l) 
\nonumber\\
&
- \frac{(p_i \cdot p_k)(p_j \cdot q)}{(p_i \cdot q)}
- \frac{(p_i \cdot p_l)(p_j \cdot q)}{(p_i \cdot q)} 
\nonumber\\
&
+ \frac{(p_j \cdot q)(p_k \cdot p_l)}{(p_i \cdot q)}
- 2 \frac{(p_i \cdot q) (p_j \cdot p_k)}{(p_j \cdot q)}
\nonumber\\
&
- 2 \frac{(p_i \cdot q)(p_j \cdot p_l)}{(p_j \cdot q)}
- 2  \frac{(p_i \cdot p_k)(p_l \cdot q)}{(p_k \cdot q)}
\nonumber\\
&
- \frac{(p_j \cdot p_k)(p_l \cdot q)}{(p_k \cdot q)}  
- 2 \frac{(p_i \cdot p_l)(p_k \cdot q)}{(p_l \cdot q)}
\nonumber\\
&
- \frac{(p_j \cdot p_l)(p_k \cdot q)}{(p_l \cdot q)}
+ \frac{(p_i \cdot p_k) (p_j \cdot q) (p_l \cdot q) }{(p_i \cdot q) (p_k \cdot q)}
\nonumber\\
&
+ \frac{(p_i \cdot p_l)(p_j \cdot q)(p_k \cdot q)}{(p_i \cdot q)(p_l \cdot q)} 
\nonumber\\
&
+ 2 \frac{(p_i \cdot q) (p_j \cdot p_k) (p_l \cdot q)}{(p_j \cdot q) (p_k \cdot q)}
\nonumber\\
&
+ 2 \frac{(p_i \cdot q)(p_j \cdot p_l)(p_k \cdot q)}{(p_j \cdot q)(p_l \cdot q)}
\biggr] \;,
\end{align}
\endgroup
where we have used four momentum conservation and the on-shell conditions
$q^2=0$ and $p_i^2 = p_j^2 = p_k^2 = p_l^2 = m_\nu^2$.
$M_\nu$ is the magnetic dipole moment of the neutrino
and $G_F$ is the Fermi constant.

\begin{figure}
\centering
\includegraphics[width=0.83\linewidth]{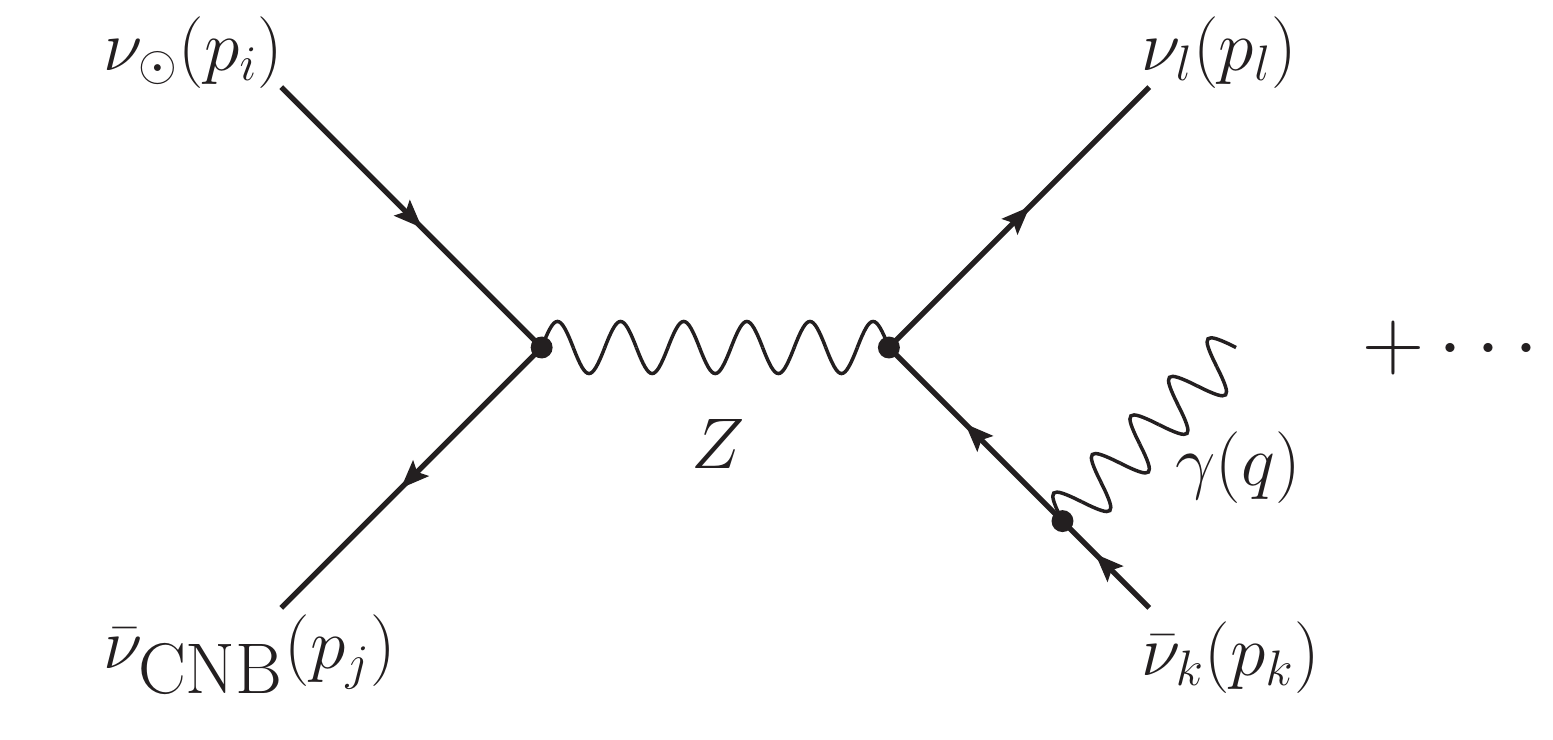}%
\caption{\label{fig:FeynmanA}
	Example Feynman diagram for the Bremsstrahlung emission from neutrino-antineutrino
	scattering. We assume in this paper that the photons couple to neutrinos
	and antineutrinos
	via a magnetic dipole moment, cf.~Eq.~\eqref{eq:vertex}.
	There are seven more diagrams. Three more diagrams where the photon
	couples to the other neutrino and anti-neutrino lines and four more
	similar to the diagram in Fig.~\ref{fig:FeynmanN}.
	The diagrams similar to the one shown in this figure would not contribute for
	different neutrino flavors in the initial state.
	}
\end{figure}

For the process involving  anti-neutrinos, cf.~Fig.~\ref{fig:FeynmanA},
\begin{equation}
	\nu_\odot(p_i) + \bar{\nu}_{\text{CNB}}(p_j)  \to  \nu (p_k) + \bar{\nu}(p_l) +  \gamma (q) \;,
\end{equation}
we find
\begingroup
\allowdisplaybreaks
\begin{align}
&\frac{|\mathcal{M} (\nu \bar{\nu} \to \nu \bar{\nu} \gamma)|^2}{128 \, G_F^2 \,  M_\nu^2} =	6 \, (p_i \cdot p_l) (p_j \cdot p_k) \nonumber\\
&+m_\nu^2 \Biggl(
2 \, (p_i \cdot p_k)
- 2 \, (p_i \cdot p_j)
+ 4 \, (p_i \cdot p_l)
\nonumber\\
&
+ 2 \, (p_j \cdot p_k)
+ (p_j \cdot p_l)
- (p_k \cdot p_l)
\nonumber\\
&
- \frac{(p_i \cdot p_j)(p_l \cdot q)}{(p_i \cdot q)} 
+ \frac{(p_i \cdot p_k)(p_l \cdot q)}{(p_i \cdot q)}
\nonumber\\
&
+ \frac{(p_j \cdot p_k)(p_l \cdot q)}{(p_i \cdot q)} 
- 2 \frac{(p_i \cdot p_j)(p_k \cdot q)}{(p_j \cdot q)}
\nonumber\\
&
+ \frac{(p_j \cdot p_l)(p_k \cdot q)}{(p_j \cdot q)} 
+ \frac{(p_i \cdot p_k)(p_j \cdot q)}{(p_k \cdot q)}
\nonumber\\
&
- \frac{(p_j \cdot q)(p_k \cdot p_l)}{(p_k \cdot q)}
+ 2 \frac{(p_i \cdot q)(p_j \cdot p_l)}{(p_l \cdot q)}
\nonumber\\
&
- 2 \frac{(p_i \cdot q)(p_k \cdot p_l)}{(p_l \cdot q)}
- \frac{(p_i \cdot p_j)(p_k \cdot q)(p_l \cdot q)}{(p_i \cdot q)(p_j \cdot q)}
\nonumber\\
&
+ \frac{(p_i \cdot p_k) (p_j \cdot q) (p_l \cdot q) }{(p_i \cdot q)(p_k \cdot q)} 
\nonumber\\
&
+ 2 \frac{(p_i \cdot q)(p_j \cdot p_l)(p_k \cdot q)}{(p_j \cdot q)(p_l \cdot q)}
\nonumber\\
&
- 2 \frac{(p_i \cdot q)(p_j \cdot q)(p_k \cdot p_l)}{(p_k \cdot q)(p_l \cdot q)}
\Biggr) \;.
\end{align}
\endgroup
In both cases, the matrix element decomposes into a part which is proportional
to the neutrino scattering case without additional photon, e.g.,
$|\mathcal{M}(\nu \nu \to \nu \nu)|^2 = 64 \, G_F^2  (p_i \cdot p_j) (p_k \cdot p_l)$,
and a part suppressed by neutrino masses.

In most cases, the matrix element is well approximated by the first term. Only in the
very forward direction for small scattering angles, the denominators in the suppressed
term can become of order $m_\nu^2$ and they become of similar importance.

With the matrix elements at hand, we can calculate the actual cross section
used in Eq.~\eqref{eq:Rate}
\begin{align}
\begin{split}
 \textrm{d}\sigma_{\nu \nu}{(E_\nu = E_i)}  &=  \frac{1}{(2 \, \pi)^5} \int \frac{\diff^3 p_k}{2 \, E_k} \frac{\diff^3 p_l}{2 \, E_l}  \frac{\diff^3 q}{2 \, E_\gamma}  \\
 &\times \frac{ |{\cal M}(\nu \nu \to \nu \nu \gamma)|^2 }{4 \sqrt{(p_i \cdot p_j)^2 - m_\nu^4 }} 
\\
 &\times \delta^4(p_i + p_j - p_k - p_l - q ) \;,
 \end{split}
\end{align}
and for $\textrm{d}\sigma_{\nu \bar{\nu}} $ we just replace the matrix element.
The calculation of the phase space is difficult analytically and we performed it numerically
at the same time with the integration over the energy spectrum of the incoming solar neutrinos
to produce the distributions in Figs.~\ref{fig:dRdEg} and \ref{fig:dRdthg}.

Actually, we can use the results for the matrix elements to provide a rough estimate for
the cross section which is supported by our numerical results. Ignoring the small terms
proportional to neutrino masses the momentum structure
is the same as for ordinary neutrino scattering which
has a cross section of the order of 
\begin{align}
 \sigma_2 \sim \frac{G_F^2}{4 \, \pi} s \;.
\end{align}
Now in our case we need to get an additional 
factor $M_\nu^2$ from the coupling to the photon which has units of inverse mass
squared. Since the momenta in the matrix elements are all taken care of
already by $\sigma_2$ we can only compensate
it by the neutrino masses. Hence, we can guess
the order of the cross section to be
\begin{align}
 \sigma \sim \frac{G_F^2}{4 \, \pi} \frac{M_\nu^2}{4 \, \pi} m_\nu^2 \, s  \approx 7.4 \times 10^{-93} \, f_M^2 \text{ cm}^2 \;.
\end{align}
To get the rate we multiply this with the total
solar neutrino flux on earth
$\Phi_{\odot} \approx 10^{11}$~cm$^{-2}$~s$^{-1}$ and
the total relic neutrino number density
$n_\nu \approx f_n \times 300$~cm$^{-3}$
to get
\begin{align}
 R \approx 7.0 \times 10^{-57} \, f_n \, f_M^2 \frac{1}{\text{km}^3 \text{ year}} \;.
\end{align}
This is actually close to our true value of
$R \approx 2.1 \times 10^{-56} \, f_n \, f_M^2$~km$^{-3}$~year$^{-1}$.

While this estimate is quite simple, we want to emphasize
that we can only trust it since it is backed by our full calculation.
For instance, if the cross section would scale with $s^2$ or $m_{\nu}^4$
instead,
we would get completely different results which justifies the
detailed calculations discussed in the main text.

Nevertheless, should a reader consider this process in a different
context the simple estimate presented here might be useful
to get a rough idea of the expected rates.

\end{document}